\documentclass[pre,superscriptaddress,showkeys,showpacs,twocolumn]{revtex4-1}
\usepackage{graphicx}
\usepackage{latexsym}
\usepackage{amsmath}
\begin {document}
\title {Language competition in a population of migrating agents}
\author{Dorota Lipowska}
\affiliation{Faculty of Modern Languages and Literature, Adam Mickiewicz University, Pozna\'{n}, Poland}
\author{Adam Lipowski}
\affiliation{Faculty of Physics, Adam Mickiewicz University, Pozna\'{n}, Poland}
\begin {abstract}
Influencing various aspects of human activity, migration is associated also with language formation. To examine the mutual interaction of these processes, we study a Naming Game with migrating agents. 
The dynamics of the model leads to  formation of low-mobility clusters, which turns out to break the symmetry of the model: although the Naming Game remains symmetric, low-mobility languages are favoured. High-mobility languages are gradually eliminated from the system and the dynamics of language formation considerably slows down. Our model is too simple to explain in detail language competition of migrating human communities, but it certainly shows that languages of settlers are favoured  over nomadic ones.

\end{abstract}

\maketitle
\section{Introduction}
The constatation that complex systems can be regarded as composed of many interacting subunits opens up a possibility of studying them with methods that were primarily developed in the context of physical many-body systems. Such  approach turned-out to be very successful~\cite{loreto}, and lead to the emergence of new research fields like  socio- or econophysics~\cite{stauffer}. Even certain linguistic problems can be studied using methods with a strong physical flavour. Language emergence~\cite{ellis} or death \cite{abrams}, its diversification~\cite{serva} and diffusion~\cite{patriarca}, time-evolving structure~\cite{petersen}, appearance of grammar or linguistic categories~\cite{puglisi}, and language learning~\cite{blythe} are just a few examples of problems, where physicists' contributions might prove to be valuable.

Of course, language, as one of our human attributes, is interrelated with many other forms of our activity. Social interactions,  economical status or political situation influence the way language is acquired and changed, or sometimes falls into oblivion. Language, as an integral part of our culture and way of life, is also intricately related to  migrations of people~\cite{routledge,williams}.
Various tribes, ethnic groups, or even entire nations firmly settled  certain areas, while some others, due to various reasons, almost constantly migrate. Migration might mix as well as separate human communities and language formation processes should be thus strongly influenced by such a factor.  Moreover, some modern trends,  especially globalization, most likely increase people's migrations \cite{czaika}. Some researches even suggest that merging multinational and multicultural migrants creates in some areas a new kind of {\it super-diverse} societies, and to describe their intercommunication, traditionally understood languages do not seem to be sufficient~\cite{jorgensen}.

It would be desirable to have some general understanding of how migration affects the language formation processes and perhaps {\it vice versa}.  As for the language formation, an interesting class of models originates from the so-called Naming Game~\cite{steels}. In this model a population of agents negotiates a language (or, more generally, conventional forms). Although the dynamics might depend on, for example, the structure of the interaction network, typically the model reaches a consensus on the language. The process of language formation resembles the ordering dynamics of Ising or Potts models accompanied, due to the symmetry of the Naming Game, by a spontanous symmetry breaking. One can even introduce the notion of an effective surface tension to explain some dynamical characteristics of the Naming Game~\cite{baron2006,lippref}. 

In the present paper, we thus examine the Naming Game in a population of migrating agents. When mobility of agents is uniform in the entire population, the model is very similar to the Naming Game of immobile agents. However, an interesting behaviour appears when the mobility depends on the language used by an agent. In such a case,  the dynamics turned out to break the symmetry of the Naming Game, favouring low-mobility languages. During the  coarsening, agents form low-mobility clusters that effectively attract and convert high-mobility neighbours. As a result, the low-mobility agents become more widespread, which considerably slows down the dynamics. Of course, our model is too simple to explain the intricacies of language competition  in settled and nomadic communities, nevertheless, it shows that the (difference in) mobility has a strong effect on such proccesses.

\section{Model}

\begin{figure}
\includegraphics[width=\columnwidth]{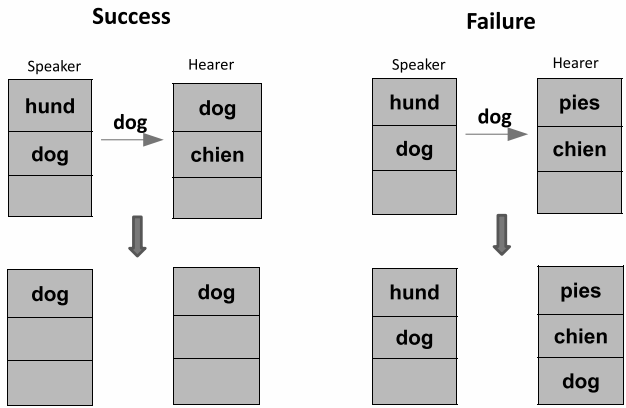}
\vspace{-8mm}
\caption{Illustration of the Naming Game dynamics. In the case of success---when the word selected by Speaker (dog) is known to Hearer---they both retain only the transmitted word in their inventories. Failure occurs when Hearer does not know the transmitted word, which is then added by Hearer to its inventory.}
\label{ng}
\end{figure}

In our model, we have a population of agents placed on a square lattice of linear size~$L$ (with periodic boundary conditions). Initially agents are uniformly distributed on the lattice with the density (i.e., probability)~$\rho$.  Each agent has its own inventory, which is a dynamically modified list of words.
The dynamics of our model combines the lattice gas diffusion with the so-called minimal version of the Naming Game~\cite{steelsbaron}.  
More specifically, in an elementary step, an agent (Speaker) and one of its neighbouring sites are randomly selected. If the selected site is empty, Speaker moves to this site. If the selected site is occupied by an agent (Hearer), then the pair Speaker-Hearer plays the Naming Game: 
\begin{itemize}
\item Speaker selects a word randomly from its inventory and transmits it to Hearer.
\item If Hearer has the transmitted word in its inventory, the interaction is a success and both players maintain only the transmitted word in their inventories.
\item If Hearer does not have the transmitted word in its inventory, the interaction is a failure and Hearer updates its inventory by adding this word to it.
\end{itemize}
The unit of time ($t=1$) is defined as $\rho L^2$ elementary steps, which corresponds to a single (on average) update of each agent. In the following, we will refer to words communicated by agents as languages. Rules of the Naming Game are also illustrated in Fig.~\ref{ng}. 

\section{Two-language version}
\begin{figure}
\includegraphics[width=\columnwidth]{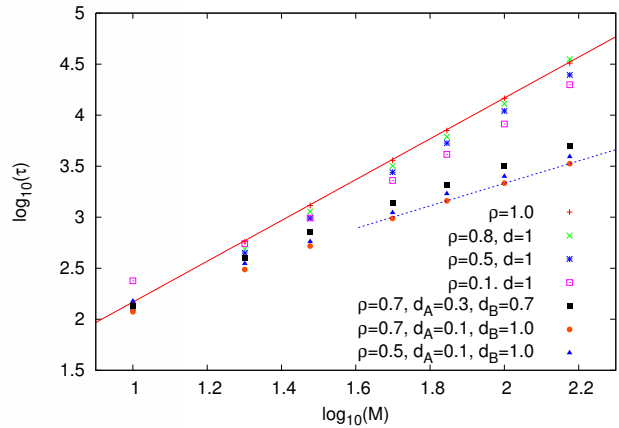}
\vspace{-8mm}
\caption{The average lifetime~$\tau$ of language~B whose users are initially in a square of size~$M$, surrounded by language~A users ($L=300$). The straight and dotted  lines have slopes corresponding to $\tau\approx M^2$ and $\tau\approx M^{1.1}$, respectively. The data for $\rho<1$ and language-independent mobility ($d_A=d_B=d=1$) also seem to obey such scaling. A different behaviour can be seen for language-specific mobility: the language~B is either quickly extinct ($d_B>d_A$) or is relatively persistent ($d_B<d_A$). Numerical results are averages over 100 independent samples.}
\label{tau_same}
\end{figure}

When $\rho=1$, all sites are occupied, thus there is no diffusion  and the model is equivalent to an ordinary square-lattice Naming Game. For $\rho<1$, a fraction $1-\rho$ of sites is empty and in addition to playing the Naming Game, agents change their locations  from time to time. There are several characteristics that might be determined for Naming Game models.
To demonstrate some analogies  to Ising-type models, we examined a two-language version of the Naming Game \cite{baron2007}. We measured the average time~$\tau$ needed for a system to reach a consensus, i.e., the state where every agent has the same language in its inventory. The initial configuration includes a square of size~$M$, inside of which all agents have  language~B in their inventories, while outside agents have  language~A. Both within the square of size~$M$ and outside, the agents are distributed with the uniform density~$\rho$. In Ising-like models, general arguments, which refer to the notion of a surface tension and the Laplace law of excesive pressure, estimate the lifetime of such a bubble as $\tau\sim M^2$~\cite{fisher}.
Our numerical results (Fig.~\ref{tau_same}) are in a very good agreement with such estimation both for $\rho=1$ and $\rho<1$ (a slight deviation for $\rho=0.1$ can be attributed to the finite size effects). It is thus a strong evidence that for migrating agents, the domain dynamics in the Naming Game is also driven by an effective surface tension. 

Let us notice that the relation $\tau\sim M^2$ is expected to hold when the bubble and its surroundings are thermodynamically equivalent phases.  In the Ising models it means that there is no external magnetic field, which would favour one of them. In the Naming Game, we have also such symmetry since the dynamics of the Naming Game does not favour any of the languages used by agents. We do not present here our additional numerical results, though we have also measured some other characteristics  of the Naming Game with migration (such as the average time needed to reach a consensus for a system initialized with  randomly assigned languages) and they qualitatively agree with the ordinary $\rho=1$ version~\cite{baron2006,lippref}.

Instead, we would like to examine an extension of the above defined model, in which to each language  its own (thus language-specific) mobility~$d$ is assigned. An agent that changes its language changes thereby also its mobility (which, we hope, might reflect the behaviour in some human communities). Now, if the chosen neighbouring site is empty,  Speaker migrates to this empty site with the probability~$d$  corresponding to the language it uses. If Speaker has several languages in its inventory, then one of them is selected randomly to determine the probability of migration (though only a very small fraction of agents have more than one language in its inventories). 

\begin{figure}
\includegraphics[width=\columnwidth]{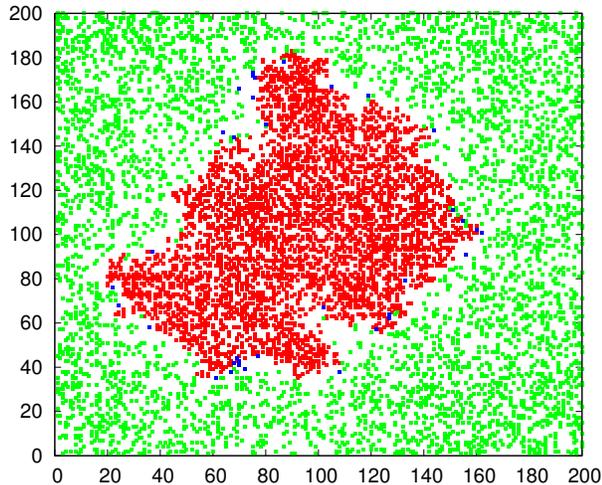}
\vspace{-5mm}
\caption{The initial configuration ($L=200$, $\rho=0.2$) includes in the center a square of size $M=30$ with B-speaking agents (red/gray),   surrounded by A-speaking agents (green/light gray). B-speakers are less mobile ($d_B=0.1$) than A-speakers ($d_A=1.0$). The figure presents a configuration after $t=5\cdot 10^4$ steps. Agents that have both~A and~B in their inventories are marked in blue/dark gray.}
\label{config}
\end{figure}

Let us notice that such modification affects only the dynamics of migration while the Naming Game remains symmetric. We determined the average time~$\tau$ in a two-language version of this model (Fig.~\ref{tau_same}). When mobility~$d_B$ of language~B users is larger than mobility~$d_A$ of surrounding language~A users, $\tau$ still increases but considerably slower and perhaps linearly $\tau \sim M$ (the least square fitting gives $\tau \sim M^{1.1}$ but a slight bending of our data makes the asymtotic $\tau \sim M$ very plausible). We do not present the estimation of time for $d_B<d_A$ since it can be made only for very small~$M$. In turn, for $M$ above a certain threshold value, the initial bubble instead of shrinking starts to grow and eventually B-users engulf the entire system. An example of such growth can be seen in Fig.~\ref{config}.

Such behaviour resembles the the behaviour of the Ising model but in the presence of an external magnetic field. When the magnetic field does not favour the bubble, a finite velocity of shrinking is expected for large~$M$ and that would explain the growth $\tau\sim M$~\cite{rikvold}. For the field favouring the bubble, the existence of a threshold size, above which the bubble will grow indefinitely, is also a well-known feature. It suggests that in our model the difference in mobility acts as a magnetic field in the Ising model and breaks the symmetry of the Naming Game favouring low-mobility languages.

\begin{figure}
\includegraphics[width=\columnwidth]{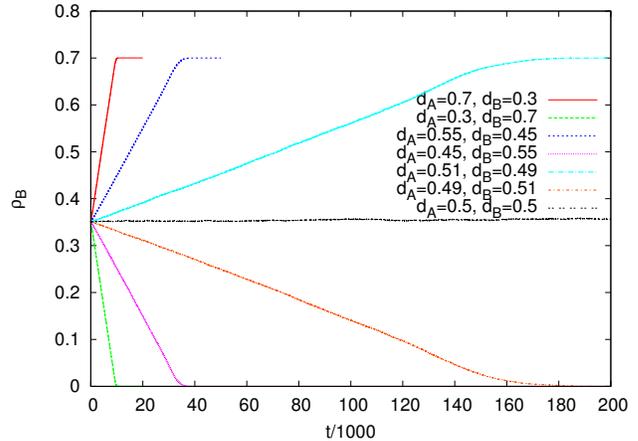}
\vspace{-7mm}
\caption{The time evolution of the density $\rho_B$ of language~B users. In the inital state, the lattice ($L=500$) is divided into two halfs containing only A- or B-speakers, respectively ($\rho=0.7$). The vertical interface thus created moves in the direction which depends on the difference in mobilities of languages. Only for equal mobilities ($d_A=d_B=0.5$), the interface remains immobile. The results are averages over 100 independent runs.}
\label{tau_half}
\end{figure}

To confirm that $d_A-d_B$ is an analogue to the magnetic field in the Ising model, we made simulations of the system initially divided (say vertically), in which the left half is filled with agents of mobility~$d_A$ and the right half with agents of mobility~$d_B$. Depending on the sign of the difference $d_A-d_B$, the interface should move (possibly at constant speed) either to the left or to the right, and only for $d_A=d_B$ it shoud stay more or less in the inital position. Our simulations fully confirmed such  scenario (Fig.~\ref{tau_half}). The  interface always moves in such a way that a less-mobile language becomes more widespread. Let us notice that even a very small difference $d_A-d_B$ is sufficient to favour one language over the other, and only for precisely the same mobilities $d_A=d_B$, the languages are equivalent.

\begin{figure}
\includegraphics[width=\columnwidth]{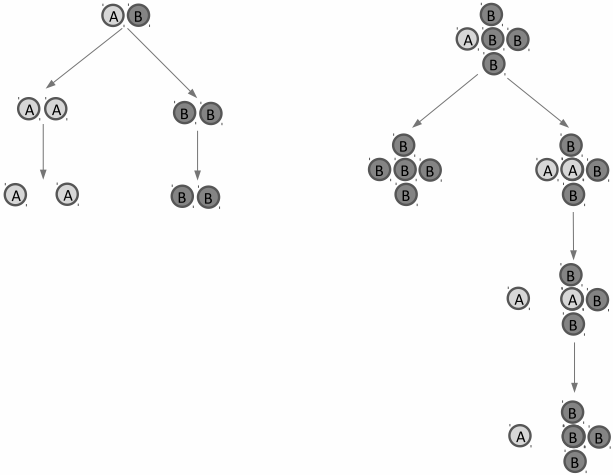}
\caption{(Left) Due to the symmetry of the Naming Game, the A-B pair with equal probabilities ends up in the A-A or B-B state. When language~A is more mobile, the pair is likely to drift apart, while the B-B pair is more stable. (Right) Apparently, the A-B symmetry is broken for more complex interactions. Even if A happens to graft his language onto a cluster, it is likely to drift appart. Then a single A~user, surrounded by three B~neighbours, is likely to be converted back to~B. Thus less mobile clusters are relatively immune to encounters with more mobile agents and the dynamics (effectively) favours less mobile languages.}
\label{processes}
\end{figure}

Let us emphasize that the rules of the Naming Game do not favour any of the languages, and the bias that appears for unequal mobilities is  generated dynamically. In our opinion, the asymmetry appears due to a tendency of low-mobility languages to form clusters. Such low-mobility clusters are relatively resistant upon interactions with high-mobility agents (Fig.~\ref{processes}). That low-mobility languages have a tendency to form clusters can be seen also in Fig.~\ref{config}. Indeed, the central area with the less mobile language seems to be more densely filled than its surroundings (and initially the entire lattice was filled with the same density $\rho=0.2$). Moreover, the interface between the languages has a considerably lower density than the interior of the area with the more mobile language. Apparently, high-mobility agents that are close to the interface get intercepted by the low-mobility center (and converted into low-mobility agents).

\begin{figure}
\includegraphics[width=\columnwidth]{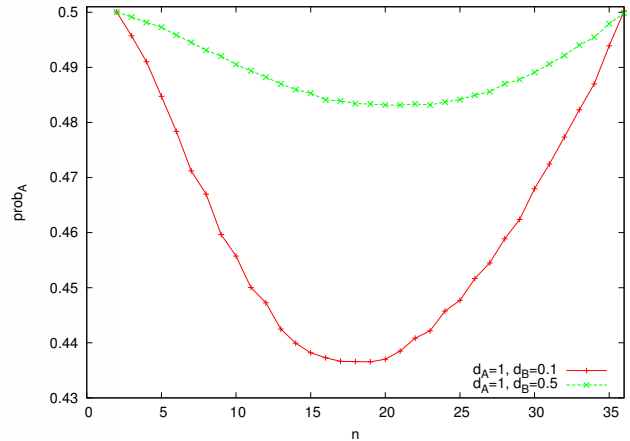}
\caption{Probability that consensus will be reached on language A ($d_A=1$) as a function of the number of agents $n$. Simulations were made for a small lattice with $L=6$. For each $n$ we made $10^6$ runs with a random distribution of agents (initially, languages A and B were also assigned randomly). Only for $n=2$ (pair interactions) and $n=36$ (no migration), the dynamics of the model remains symmetric and the probability to end up with language~A is the same as with language~B.}
\label{prob6x6}
\end{figure}

To support the above arguments, we made simulations  where we measured the probability prob$_A$ that a small system ($L=6$) starting with randomly distributed agents will reach a consensus with  language A. Initially, languages A and B (and migrations $d_A$ and $d_B$) are also randomly assigned to agents. Numerical simulations show (Fig.~\ref{prob6x6}) that only for the number of agents $n=2$ and $n=36$, we have prob$_A=0.5$. For $n=2$, only binary interactions of agents might take place (left panel of Fig.~\ref{processes}) and the symmetry of the Naming Game implies that prob$_A=0.5$. Similarly, for $n=36$ migration is suppressed and prob$_A=0.5$ is the expected ordinary Naming Game result. Simulations show, however, that for any other value of $n$ the symmetry of the model is broken and the less mobile language (B) is effectively favoured. 
It would be certainly desirable to have more general understanding of the mobility-induced symmetry breaking that takes place in our model. For example, one might hope to develop some kind of a coarse-grained description of our model in terms of the Ginzburg-Landau potential, an approach that turned out to be quite effective in some other agreement-dynamics models \cite{dallasta2008,vazquez}.

\section{Multi-language version}
\begin{figure}
\includegraphics[clip,width=\columnwidth]{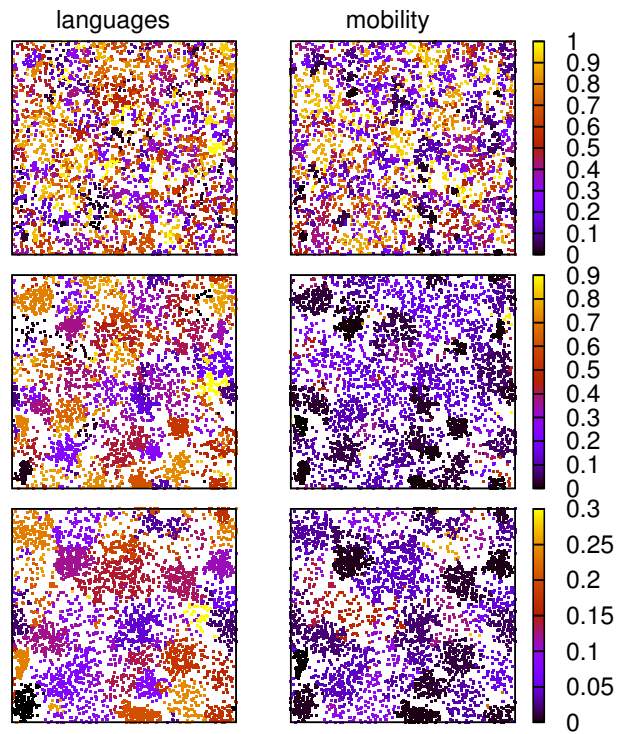}
\vspace{-8mm}
\caption{The configuration obtained after $t=5\cdot 10^2$ (top), $3\cdot 10^3$ (middle) and $t=10^4$ (bottom) MC steps (\mbox{$L=200$}, \mbox{$\rho=0.1$}).  Low-mobility languages form more dense clusters, surrounded by depleted zones, and gradually grow at the expense of more mobile neighbours. The color bars apply only to the mobility panels.}  
\label{n-config}
\end{figure}

In the present section we examine the multi-language version of our model.
In such a case Fig.~\ref{n-config} shows that the cluster-formation mechanism is also at work (simulations start from a random distribution of languages and their mobilities). Initially mobilities were set randomly from the range $0<d<1$ but by the time $t=3\cdot 10^3$ and especially $t=10^4$, languages with the largest mobilities (close to~1) were eliminated. One can clearly see the formation of low-mobility clusters, which grow by depleting their surroundings from more mobile agents. 

\begin{figure}
\includegraphics[width=\columnwidth]{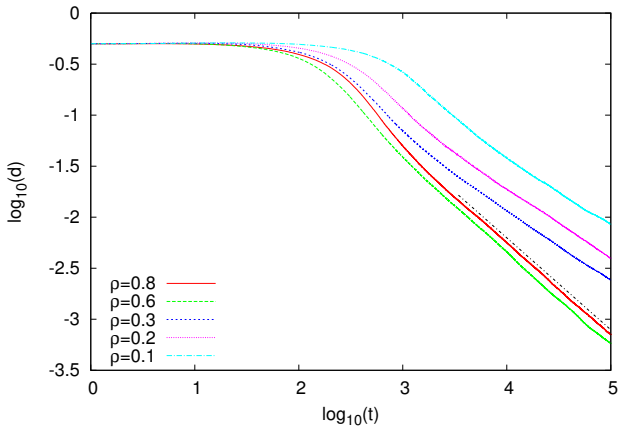}
\vspace{-8mm}
\caption{The time dependence of the average mobility~$\langle d \rangle$; simulation for $L=10^3$ with random initial conditions and averaged over 100 independent runs. The dash-dotted line corresponds to the decay $\langle d \rangle\sim t^{-0.9}$.}
\label{mobility-decay}
\end{figure}

We also examined the time dependence of the average mobility~$\langle d \rangle$ in the system. Indeed, the numerical results in Fig.~\ref{mobility-decay} confirm that  $\langle d \rangle$ systematically decreases. For a low density ($\rho=0.1$), one can notice relatively long initial plateaux, related to the fact that the system needs some time to build low-mobility clusters, and only then the process that favours low-mobility languages starts.
Moreover, in the time decay of the average mobility, one can distinguish two power-law regimes. In the high-density regime ($\rho=0.6$ and 0.8), one has $\langle d \rangle\sim t^{-0.9}$ while in the low-density regime ($\rho=0.1,$ 0.2, and 0.3), the exponent is smaller than 0.9 and perhaps even varies with density. 

\begin{figure}
\includegraphics[width=\columnwidth]{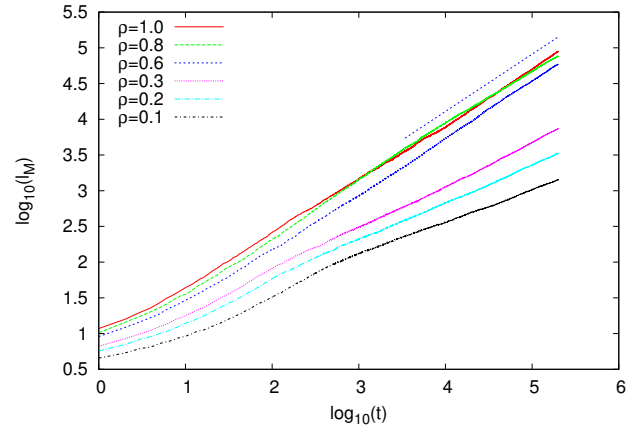}
\vspace{-8mm}
\caption{The time dependence of the number of users of the most common language~$l_M$; 
simulation for $L=10^3$ with random initial conditions and averaged over 100 independent runs.
The dashed line corresponds to the increase $l_M\sim t^{0.8}$.}
\label{max_lang}
\end{figure}

Certain Naming Game characteristics exhibit a similar power-law behaviour. In Fig.~\ref{max_lang} we present the time dependence of the number of users of the largest language~$l_M$. While in the high-density regime ($\rho=1.0$, 0.8, 0.6), the increase seems to be universal and $l_M\sim t^{0.8}$, in the low-density regime ($\rho=0.3$, 0.2, 0.1), the power-law behaviour has a density dependent exponent. The behaviour of~$\langle d \rangle$ and~$l_M$ shows that the dynamics in the high-density regime is much faster and the Naming Game characteristics are very similar to those of an ordinary Naming Game (with $\rho=1$). The low-density regime has a much slower dynamics and we relate such  behaviour to the formation of low-mobility clusters (Fig.~\ref{n-config}). It is likely that the model undergoes a phase transition around $\rho=0.5$, but its more detailed analysis is left for the future.

The two-language version described in the previous section exhibits a symmetry breaking that speeds up the dynamics.
In the multi-language version, however, we have initially the entire spectrum of languages and mobilities. The dynamics gradually eliminates large-mobility agents  and thus the remaining small-mobility agents are primarily responsible for a considerably slower dynamics (Fig.~\ref{max_lang}).

\section{Conclusions}
In summary, motivated by a possible mutual influence of language formation and migration of human communities, we examined the Naming Game model with mobile agents. As our main result, we have shown that even a small difference in a language-specific mobility favours a low-mobility language. 
Of course, taking into account an extreme complexity of human interactions, we are not even tempted to suggest that our model proves that languages of settlers should outperform nomadic ones, nevertheless, it certainly shows a strong relation between language formation and migration.
In our model, low-mobility languages form clusters and in a low-density regime this process slows down the dynamics of the Naming Game. Let us also notice that  the dynamics of a typical Naming Game (with $\rho=1$) rather quickly leads to the consensus, which not necessarily corresponds with a relatively stable multi-language structure of the human population~\cite{lipweighted}. With this respect, a slower dynamics and a longer lifetime of the multi-language state (as suggested in Fig.~\ref{max_lang}) of the proposed model might be more suitable. 
Finally, it should be also noted that the Naming Game is one of the models with the so-called agreement dynamics. The Voter or Ising models are yet other well-known examples of this kind of models and some aspects of mobility in such systems were already examined \cite{baronsatorras}. It would be, in our opinion, interesting to examine their generalizations that taking into account the state-dependent mobility.



\end {document}